\documentclass[12pt,notitlepage,aps,amsmath,amssymb]{revtex4-1}

\usepackage{bm}
\usepackage{mathtools}
\usepackage{graphicx}
\usepackage{caption,subcaption}
\usepackage{amsthm}

\usepackage{times}
\usepackage{newtxtext}

\newcommand{\Eqref}[1]{Eq.~\eqref{#1}}

\newcommand{\Sn}{\Phi}
\newcommand{\nSn}{\widetilde{\Sn}}
\newcommand{\dT}{{\triangle T}}
\newcommand{\sn}{\phi}
\newcommand{\tw}{\tau_\text{w}}
\newcommand{\td}{\tau_\text{d}}

\newcommand{\m}[1]{\left< #1 \right>} 
\newcommand{\tm}[1]{\langle #1 \rangle} 
\newcommand{\wh}[1]{\widehat{#1}}
\newcommand{\mA}{\m{A}}
\newcommand{\rmi}{\text{i}}
\newcommand{\rmd}{\text{d}}
\newcommand{\rms}{\text{rms}}

\let\originalleft\left
\let\originalright\right
\renewcommand{\left}{\mathopen{}\mathclose\bgroup\originalleft}
\renewcommand{\right}{\aftergroup\egroup\originalright}

\begin{document}

\title{Scaling of shot noise processes}

\author{A.~Theodorsen}
\email{audun.theodorsen@uit.no}
\affiliation{Department of Physics and Technology, UiT The Arctic University of Norway, N-9037 Troms{\o}, Norway}

\date{\today}

\begin{abstract}
   In this contribution, we investigate the scaling of the distribution of the shot noise process, its power spectral density and its time above threshold.
\end{abstract}

\maketitle

The shot noise process is given by
\begin{equation}
	\Sn(t) = \sum_{k=0}^\infty A_k \sn\left(\frac{t-t_k}{\tau_k}\right).
\end{equation}
We assume exponential waiting times and uncorrelated random variables.
We have that the characteristic function of $\Sn$ is given by \cite{theodorsen-ppcf}
\begin{equation}
\ln C_\Sn(u) = \gamma \sum_{n=1}^\infty  \frac{(\rmi u)^n}{n!} \m{A^n} I_n
\end{equation}
where $\gamma = \tm{\tau}/\tw$ and $I_n = \int_{-\infty}^{\infty} \sn(\theta)^n \rmd \theta$.

The power spectral density of $\nSn = (\Sn - \tm{\Sn})/\Sn_{\rms}$ is given by \cite{garcia-pop-1}
\begin{equation}\label{eq:psd}
\Omega_{\nSn}(\omega) = \frac{1}{\td} \int\limits_0^\infty \rmd \tau \, \tau^2 P_\tau(\tau) \varphi(\omega \tau).
\end{equation}

The general method for deriving the time above threshold in the limit of no pulse overlap is given in \cite{theodorsen-pre}. In the normal limit, the shot noise process $\nSn$ is a process with standard normal distribution and power spectrum given by \Eqref{eq:psd}. If the power spectral density scales as a power law, $\Omega \sim \omega^{-\beta}$, the shot noise process in the normal limit will follow a fractional Brownian motion (fBM) with Hurst parameter given by $\beta = 2H+1$, $0<H<1$. It is known that the first return time for fBM (which is equivalent to time above threshold) scales as $\dT^{H-2}$ \cite{rypdal-pre}. Thus, in the normal limit we straightforwardly have $\nu = (5-\beta)/2$. In the same manner, it can be shown that the mass above threshold $S = A\int_0^{\dT} \phi(t/\td)\rmd t$ scales as $p_S(S) \sim S^{-\chi}$ with $\chi = 2/(1+H) = 4/(1+\beta)$.

Unless indicated, duration times are assumed to be degenerately distributed, pulses are exponential functions and the amplitudes are exponentially distributed.

\begin{centering}
\begin{tabular}{ c | c| c |  c  | c  }
& Standard & $p_\tau(\tau)\sim \tau^{-\alpha}$ & $\sn(\theta) \sim \theta^{-\alpha}$ & $p_A(A) \sim A^{-\alpha}$ \\
\hline 
regime & & $1<\alpha<3$ & $0<\alpha<1$ & $1<\alpha<3$ (?) \\
\hline
$p_{\Sn}(\Sn) \sim \Sn^{-s}$ & None & None & None & Present \\
\hline
$\Omega_{\nSn}(\omega) \sim \omega^{-\beta}$ & None (0, 2) & $\beta = 3-\alpha$ & $\beta = 2(1-\alpha)$ & None \\
\hline
Intermittent limit $p_\dT(\dT) \sim \dT^{-\nu}$  & None & $\nu = \alpha$ & None ($\nu = 1-\alpha$ )& None\\
\hline
Normal limit $p_\dT(\dT) \sim \dT^{-\nu}$ & $\nu = 3/2$  & $\nu = \alpha/2 +1 $  & $\nu = \alpha+3/2$ & $\nu = 3/2$\\
\hline
Intermittent limit $p_S(S) \sim S^{-\chi}$   & None & $\chi = \alpha$ & None & $\chi = \alpha$\\
\hline
Normal limit $p_S(S) \sim S^{-\chi}$  (?) & $\chi = 4/3$ & $\chi = 4/(4-\alpha)$ & $\chi = 4/(3-2\alpha)$ & $\chi = 4/3$ \\
\end{tabular}
\end{centering}

\section{Standard shot noise}
\paragraph{Probability distribution}
In this case, the probability distribution is known to be a Gamma distribution with no power-law scaling.

\paragraph{Power spectral density}
This has been treated in previous publications. The scaling is 0 for low values and 2 for large values, but this is the power spectrum of exponential decay.

\paragraph{Duration above threshold - intermittent limit}
In \cite{theodorsen-pre}, this is shown to be a Gumbel distribution which lacks power law scaling.

\paragraph{Duration above threshold - normal limit}
This is a known result for the Ornstein-Uhlenbeck process.

\paragraph{Mass above threshold - intermittent limit}
With the exponential pulse, $\dT = \td \ln(A/L)$, giving $S = \td (A-L)$. This means that $A>L \rightarrow S>0$. We then have
\begin{equation}\label{eq:mass-base}
p_S(S) = \frac{1}{\td} p_A\left(\frac{S}{\td} + L |  \frac{S}{\td} + L > L\right) = \frac{1}{\td \mA} \exp\left( - \frac{S}{\td \mA} \right), \, S>0.
\end{equation}
There is no scaling here.
\paragraph{Mass above threshold- normal limit}
This is a known result for the Ornstein-Uhlenbeck process \cite{kearney-2005}.

\section{Power law pulse duration times}
\paragraph{Probability distribution}
The probability density function of $\Sn$ only depends on $\tm{\tau}$, and so does not scale with any scaling parameter of $\tau$. 

\paragraph{Power spectral density}
From \Eqref{eq:psd}, the scaling of the power spectrum is straightforward. It has been seen that this only holds for $1<\alpha<3$.

\paragraph{Duration above threshold - intermittent limit}
Assuming exponential pulses and exponentially distributed amplitudes, $p_{\dT}(\dT | \tau)$ is given in \cite{theodorsen-pre} in the strongly intermittent limit. By inspection, $p_{\dT}(\lambda \dT | \tau) = \frac{1}{\lambda} p_{\dT}(\dT | \tau/\lambda)$. If $p_\tau \sim \tau^{-\alpha}$, we then have that $p_\tau(\tau) = \lambda^{-\alpha} p_\tau(\tau/\lambda)$, and
\begin{align*}
p_{\dT} (\dT) &= \int\limits_0^\infty \rmd \tau \, p_\tau(\tau) p_{\dT}(\dT | \tau) \\
p_{\dT} (\lambda \dT) &= \int\limits_0^\infty \rmd \tau \, p_\tau(\tau) p_{\dT}( \lambda \dT | \tau) \\
 				&= \int\limits_0^\infty \rmd \frac{\tau}{\lambda} \, p_\tau(\tau) p_{\dT}(\dT | \tau/\lambda) \\
				&= \lambda^{-\alpha} \int\limits_0^\infty \rmd \frac{\tau}{\lambda} \, p_\tau(\tau/\lambda) p_{\dT}(\dT | \tau/\lambda) \\
				&= \lambda^{-\alpha} p_{\dT}(\dT).
\end{align*}

\paragraph{Duration above threshold - normal limit}
Calculated from $\nu = (5-\beta)/2$.

\paragraph{Mass above threshold - intermittent limit}
Since $p_S(S | \tau)$ is an exponential distribution with mean value $\tau \mA$, we have that $p_S(\lambda S | \tau) = \frac{1}{\lambda} p_S(S | \tau/\lambda)$, the calculation follows analogously to the one above and we have
\begin{equation}
p_S(\lambda S) = \lambda^{-\alpha} p_S(S),
\end{equation}
giving the scaling $p_S(S) \sim S^{-\alpha}$.
\paragraph{Mass above threshold- normal limit}
The scaling is derived from the expression for known $\beta$.

\subsection{The rate parameter $\lambda$}
Note that several authors have used the rate parameter $1/\tau$ instead of $\tau$. For $p_\tau \sim \tau^{-\alpha}$, $p_\lambda \sim \lambda^{\alpha-2}$. Thus at $\alpha = 1$, these are equal, but in general there is a shift in the distribution. Thus, uniform $\lambda$, which is known to give $\Omega \sim \omega^{-1}$ requires $\tau^{-2}$. Several authors have wrongly assumed $\tau^{-1}$ leads to  $\Omega \sim \omega^{-1}$ .

\section{Power law pulses}
\paragraph{Probability distribution}
Possible presence by inspection of probability distribution.
\paragraph{Power spectral density}
The scaling of $\Omega$ is given by \cite{lowen-teich}. Note that this requires $0<\alpha<1$. 
\paragraph{Duration above threshold - intermittent limit}
We consider the pulse shape
\begin{equation}
\phi(\theta) = c (\theta+m)^{-\alpha},\, 0\leq\theta\leq \Delta,
\end{equation}
where $c = [1-\alpha]/[(\Delta + m)^{1-\alpha} - m^{1-\alpha}]$ is a normalization constant. For the pulse to trigger, we require $A \phi(0) >L$. If $A \phi(\Delta) > L$, the duration is the full pulse duration, $\dT = \td \Delta$. The possible interesting scaling therefore happens for $L/\phi(0) < A < L/\phi(\Delta)$. The truncated exponential distribution for $A$ is therefore
\begin{equation}
p_A(A | L/\phi(0) < A < L/\phi(\Delta)) = \frac{1}{\mA} \frac{ \exp(-A/\mA) }{ \exp\left(- L/\mA \phi(0) \right) -  \exp\left(-L/\mA \phi(\Delta) \right)  }.
\end{equation}
We then have that (RECHECK, POSSIBLY $\dT/\td \rightarrow \dT/\td + m$!)
\begin{equation}
p_{\dT}(\dT | 0< \dT < \td \Delta) = \frac{\alpha L}{c} \frac{\dT^{\alpha-1}}{\td^\alpha} p_A\left[\frac{L}{c} \left( \frac{\dT}{\td} \right)^\alpha | L/\phi(0) < A < L/\phi(\Delta)) \right]
\end{equation}
giving
\begin{equation}
p_{\dT}(\dT | 0< \dT < \td \Delta) \propto \dT^{\alpha-1} \exp\left( -\frac{L}{c \mA} \frac{\dT^\alpha}{\td^\alpha} \right).
\end{equation}
We note that the scaling $\alpha = 0 \rightarrow p_{\dT} \sim \dT^{-1}$ is not possible, as in this case $\phi(0) = \phi(\Delta)$ and there are no possible events. For small values of the exponent (or in the limit $\dT \rightarrow 0$), the exponential function approaches 1 and $p_{\dT} \sim \dT^{\alpha-1}$.
\paragraph{Duration above threshold - normal limit}
Calculated from $\nu = (5-\beta)/2$.

\paragraph{Mass above threshold - intermittent limit}
The pulse is the same as above. For $L/\phi(0) < A < L/\phi(\Delta)$, we have $\dT = \td (A c / L)^{1/\alpha} - \td m$, and 
\begin{equation}
S = \frac{ c \td}{1-\alpha} A \left[ \left( \frac{A c}{L} \right)^{-1+1/\alpha} - m^{1-\alpha} \right].
\end{equation}
We can rewrite this equation as (using that $ X = A/\mA$ is a standard exponential distribution):
\begin{align}
\frac{1-\alpha}{c \td \mA} \left( \frac{L}{c \mA} \right)^{-1+1/\alpha} S &= X^{1/\alpha} - \left( \frac{L}{c \mA} \right)^{-1+1/\alpha} m^{1-\alpha} X \nonumber \\
\wh{S} &= X^{1/\alpha} - c_1 X,
\end{align}
Note that $L/\phi(0) < A  \rightarrow c_1^{\alpha/(1-\alpha)} < X$ is equivalent to $S>0$. On the other side, we require $X<c_1^{\alpha/(1-\alpha)} (1+\Delta/m)^\alpha$. Above this, $S = A \td \Delta$.
By visual inspection of numerically generated PDFs, there is no power law scaling in this function.

\paragraph{Mass above threshold- normal limit}
The scaling is derived from the expression for known $\beta$.

\section{Power law amplitudes}
\paragraph{Probability distribution}
Present by inspection of probability distribution
\paragraph{Power spectral density}
The pulse amplitude distribution does not affect the power spectral density.
\paragraph{Duration above threshold - intermittent limit}
Following \cite{theodorsen-pre}, we find that for a truncated Pareto distribution,
\begin{equation}
p_A(A) = \frac{1-\alpha}{M^{1-\alpha} - m^{1-\alpha}} A^{-\alpha},\, m<A<M,
\end{equation}
the conditional amplitude distribution for amplitudes above threshold are
\begin{equation}
p_A(A | A>L) = c A^{-\alpha},\, m<A<M,
\end{equation}
where 
\begin{equation}
c = \begin{cases}  \frac{1-\alpha}{M^{1-\alpha} - m^{1-\alpha}}, & L<m,  \\  \frac{1-\alpha}{M^{1-\alpha} - L^{1-\alpha}} , & m<L<M, \\ 0, & L>M. \end{cases}
\end{equation}
With the exponential pulse, we get that
\begin{equation}
p_{\dT}(\dT) = \frac{L}{\td} \exp\left( \frac{\dT}{\td} \right) p_A( L \exp(\dT/\td) | A>L) = c \frac{L^{1-\alpha}}{\td} \exp\left[ (1-\alpha) \frac{\dT}{\td} \right]
\end{equation}
which has no power law scaling in $\alpha$.

\paragraph{Duration above threshold - normal limit}
In the normal limit, we have a process with normal probability distribution and exponential correlation (OU-process). This gives a scaling as 1/2.

\paragraph{Mass above threshold - intermittent limit}
In this case, we still have $S = \td(A-L)$ from the calculation preceding \Eqref{eq:mass-base}. It follows straightforwardly that $p_A \sim A^{-\alpha} \rightarrow p_S \sim S^{-\alpha}$.
\paragraph{Mass above threshold- normal limit}
This is a known result for the Ornstein-Uhlenbeck process \cite{kearney-2005}.

\end{document}